\documentclass[aps,pre,preprint,superscriptaddress]{revtex4}
\usepackage{graphicx}
\begin{document}

\title{Discretization-related issues in the KPZ equation:
Consistency, Galilean-invariance violation,
and fluctuation--dissipation relation}

\author{H. S. Wio}
\affiliation{Instituto de F\'{\i}sica de Cantabria (UC and CSIC),\\
Avda.\ de los Castros, s/n, E-39005 Santander, Spain.}

\author{J. A. Revelli}
\affiliation{Instituto de F\'{\i}sica de Cantabria (UC and CSIC),\\
Avda.\ de los Castros, s/n, E-39005 Santander, Spain.}

\author{R. R. Deza}
\affiliation{Instituto de Investigaciones F\'{\i}sicas Mar del Plata
(UNMdP and CONICET),\\
De\'an Funes 3350, B7602AYL Mar del Plata, Argentina.}

\author{C. Escudero}
\affiliation{ICMAT (CSIC-UAM-UC3M-UCM), Departamento de Matem\'aticas,
Facultad de Ciencias, Universidad Aut\'onoma de Madrid,\\
Ciudad Universitaria de Cantoblanco, 28049 Madrid, Spain.}

\author{M. S. de La Lama}
\affiliation{Instituto de F\'{\i}sica de Cantabria (UC and CSIC),\\
Avda.\ de los Castros, s/n, E-39005 Santander, Spain.}
\date{\today}

\begin{abstract}
In order to perform numerical simulations of the KPZ equation, in any
dimensionality, a spatial discretization scheme must be prescribed.
The known fact that the KPZ equation can be obtained as a result of a
Hopf--Cole transformation applied to a diffusion equation (with
\emph{multiplicative} noise) is shown here to strongly restrict the
arbitrariness in the choice of spatial discretization schemes. On one
hand, the discretization prescriptions for the Laplacian and the
nonlinear (KPZ) term cannot be independently chosen. On the other
hand, since the discretization is an operation performed on
\emph{space} and the Hopf--Cole transformation is \emph{local} both in
space and time, the former should be the same regardless of the field
to which it is applied. It is shown that whereas some discretization
schemes pass both consistency tests, known examples in the literature
do not. The requirement of consistency for the discretization of
Lyapunov functionals is argued to be a natural and safe starting point
in choosing spatial discretization schemes. We also analyze the
relation between real-space and pseudo-spectral discrete
representations. In addition we discuss the relevance of the Galilean
invariance violation in these consistent discretization schemes, and
the alleged conflict of standard discretization with the
fluctuation--dissipation theorem, peculiar of 1D.
\end{abstract}
\maketitle

\section{Introduction}\label{sec:1}

Soon after its formulation in 1986, the KPZ equation
\cite{kpz,HHZ,BarSta}
\begin{eqnarray}\label{eq:1}
\partial_th=\nu\,\partial_x^2h+\frac{\lambda}{2}
\left(\partial_xh\right)^2+F+\varepsilon\,\xi(x,t),
\end{eqnarray}
became a paradigm as the description of a vast class of
nonequilibrium phenomena by means of stochastic fields. The field
\(h(x,t)\) whose evolution is governed by this stochastic nonlinear
partial differential equation, describes the height of a fluctuating
interface in the context of surface growth processes in which it was
originally formulated. In particular, Eq.\ (\ref{eq:1}) assumes a
one-dimensional (1D) homogeneous substrate of size \(L\). The
parameter \(\nu\) determines the surface tension, \(\lambda\) is
proportional to the average growth velocity (the surface slope is
parallel-transported in the growth process), and \(F\) is an
external driving force. Finally, \(\xi(x,t)\) is a Gaussian white
noise with \(\langle\xi(x,t)\rangle=0\) and
\(\langle\xi(x,t)\xi(x',t')\rangle=2\delta(x-x')\delta(t-t')\). As
usual, periodic boundary conditions are assumed.

From a theoretical point of view the KPZ equation has many
interesting properties, like its close relationship with the Burgers
equation \cite{Fogedby} or with a diffusion equation with
multiplicative noise, whose field \(\phi(x,t)\) can be interpreted
as the restricted partition function of the directed polymer
problem. But clearly, investigating the behavior of its solutions to
obtain e.g.\ the critical exponents in one or more spatial
dimensions
\cite{FoTo,BecCur,MoserWolf-Discr3d,Scalerandi-etal,NewmanBray-Discr,LamShin-Discretiz-1,LamShin-Discretiz-2,Appert,Marinari-etal}
requires the (stochastic) numerical integration of a discrete
version. Although a pseudo-spectral spatial discretization scheme
has been recently put forward
\cite{GiaRo01,GiadaGiacomettiRossi,GCL1}, as well as a numerical
large deviation theory \cite{Fogedby}, real-space discrete versions
of Eq.\ (\ref{eq:1}) are still largely used for numerical
simulations \cite{Tabei-etal,Reis,Buceta-Discr,Ma-etal-Discr2d},
because of their relative ease of implementation and of
interpretation in the case of non-homogeneous substrates (for
instance, a quenched impurity distribution \cite{delala}) among
other reasons. To that end, several real-space discretization
schemes have been proposed
\cite{LamShin-Discretiz-1,LamShin-Discretiz-2,Buceta-Discr}, which
are claimed to cure particular ``diseases'' of the numerical
simulation.

In the present work, no attempt is made of comparing alternative
real-space discretization schemes in sought of special KPZ features.
Instead, we seek to point out some basic conditions that \emph{any}
spatial discretization must fulfill in order to consistently
describe the KPZ equation. Nonetheless, for the sake of brevity and
for ease of comparison with other proposals, we shall adopt the
notation in Ref.\ \cite{Buceta-Discr}, namely (calling \(\Delta
x\equiv a\))
\begin{eqnarray*}
& & L=aN,\qquad\qquad H_{j+k}^{j+l}\equiv\frac{h_{j+l}-h_{j+k}}{a},\\
& &
L_j\equiv\frac{H_j^{j+1}-H_{j-1}^j}{a}=\frac{H_j^{j+1}+H_j^{j-1}}{a},
\end{eqnarray*}
and
\[N_j^{(\gamma)}\equiv\frac{\left(H_j^{j+1}\right)^2+2\gamma
H_j^{j+1}H_{j-1}^j+\left(H_{j-1}^j\right)^2}{2(1+\gamma)},\]
with \(\gamma\in[0,1]\). On one hand, the restriction to
\(k,l\in\{-1,0,1\}\) is unnecessary. On the other hand, we shall
denote \(L_j\rightarrow L_{(1)}(h_j)\) and
\(N_j^{(\gamma)}\rightarrow N^{(\gamma)}(h_j)\). [The subscript
\((1)\) indicates that only nearest neighbors are involved in the
prescription of the discrete Laplacian.] By analogy, we shall write
\[\Phi_{j+k}^{j+l}\equiv(\phi_{j+l}-\phi_{j+k})/a,\]
and consequently
\[L_{(1)}(\phi_j)\equiv\frac{\Phi_j^{j+1}+\Phi_j^{j-1}}{a},\]
and
\[N^{(\gamma)}(\phi_j)\equiv\frac{\left(\Phi_j^{j+1}\right)^2+2\gamma
\Phi_j^{j+1}\Phi_{j-1}^j+\left(\Phi_{j-1}^j\right)^2}{2(1+\gamma)}.\]

Two main symmetries are usually ascribed to the 1D KPZ equation:
Galilean invariance and the fluctuation--dissipation relation.
\begin{itemize}
\item The first one has been traditionally linked to the
exactness (in any spatial dimensionality) of the relation
\(\alpha+z=2\) among the roughness \(\alpha\) and dynamic \(z\)
exponents \cite{FoNeSt,mhkz89}, although this interpretation has
been recently criticized in other nonequilibrium models
\cite{BeHo,BeHo2}. The roughness exponent \(\alpha\) characterizes
the surface morphology in the stationary regime \(t \gg t_x\). On
the other hand, the correlation length scales as \(\xi(t)\sim
t^{1/z}\) with the dynamic exponent \(z\), and \(t_x\) is the time
at which it saturates, namely \(\xi(t \gg t_x)\sim L\). The ratio
\(\beta=\alpha/z\) is called ``growth exponent'' and characterizes
the short-time behavior of the interface.
\item The second symmetry essentially tells us that in 1D, the
nonlinear (KPZ) term is not operative at long times or in other
words, that the long-time 1D interface is equivalent to a path of
Brownian motion \cite{BarSta}. From a theorem by Kolgomorov, this
implies that the interface is H\"older continuous with exponent
strictly smaller than \(1/2\). For higher dimensions, the KPZ
roughness exponent \(\alpha\) decreases, implying a loss of
regularity. Hence the error terms of a local numerical method (as
e.g.\ a finite differences scheme), which are proportional to some
higher-order derivative of the field, are not controlled. As a
consequence, a global method such as a pseudo-spectral scheme
\cite{GiaRo01,GiadaGiacomettiRossi,GCL1,GiaRos} is more adequate.
Nevertheless, previous experiences found in the literature showed
that finite differences schemes are still able to capture the
universal features of KPZ evolution. This, together with our
previous considerations, is our motivation for the present work.
\end{itemize}

In Sec.\ \ref{sec:2} we show that the relationship established by
the Hopf--Cole transformation---between the KPZ equation and a
diffusion equation with multiplicative noise \cite{HHZ}---poses
constraints on the discretization procedure. We verify the
consistency of the standard (nearest-neighbor) discretization scheme
and find the form of the corresponding KPZ term for a general
real-space discrete Laplacian; we also present some comments
regarding the mapping of KPZ into the directed polymer problem
\cite{HHZ}. In Sec.\ \ref{sec:3} we analyze the problem from the
perspective of the Lyapunov functional, show in what sense known
prescriptions for the KPZ term from the literature fail the test,
and find the corresponding consistent prescriptions. Moreover, we
propose a consistent real-space discretization scheme whose accuracy
is far higher than that of schemes of similar complexity in the
literature. In Sec.\ \ref{sec:4} we discuss the relation with the
pseudo-spectral method. In Sec.\ \ref{sec:5}, we show that a
consistent discretization scheme does not (essentially) violate the
fluctuation--dissipation relation, peculiar of 1D, and discuss the
role of the Galilean invariance for the discrete representations of
the KPZ equation, showing that such invariance seems not to be a
necessary element to define the KPZ universality class. In Sec.\
\ref{sec:6} we discuss a recently introduced variational approach
for the KPZ equation \cite{Wio}, and show that it offers a natural
framework for its consistent discretization. In Section \ref{sec:7}
we present some numerical results regarding critical exponents and
the violation of Galilean invariance. Section \ref{sec:8} contains
the conclusions and final discussions. It is worth here commenting
that some preliminary results were presented in \cite{nos-KPZ}.

\section{The Laplacian determines the nonlinear term}\label{sec:2}

In this section we elucidate---by considering the standard,
nearest-neighbor discretization prescription as a benchmark---one of
two constraints to be obeyed by any spatial discretization scheme.
It is very important to remark that this constraint arises due to
the mapping between the KPZ and the diffusion equation (with
multiplicative noise) through the Hopf--Cole transformation. Hence,
for a general real-space discrete Laplacian, we state the form of
its corresponding KPZ term. Even though the present analysis is
performed on the KPZ  equation, it is general in the sense that for
sets of equations related among themselves through a local
transformation there should be a consistent relation between the
discrete transformed forms.

\subsection{The simplest case}\label{ssec:2.1}

As it is known, the diffusion equation with multiplicative noise
\begin{eqnarray}\label{eq:2}
\partial_t \phi = \nu\,\partial_x^2\phi+\frac{\lambda F}{2\nu}\phi
+\frac{\lambda\varepsilon}{2\nu}\phi\,\xi,
\end{eqnarray}
is related to the KPZ equation [Eq.\ (\ref{eq:1})] through the
\emph{Hopf--Cole transformation}
\begin{equation}\label{eq:3}
\phi(x,t)=\exp\left[\frac{\lambda}{2\nu}h(x,t)\right].
\end{equation}
Note that this transformation is just one particular example
of the \emph{general implicit transformation} written down in
Ref.\ \cite{schlbr79}.

The standard spatial discrete version of Eq.\ (\ref{eq:2}), after
transforming to a co-moving reference frame \(\phi\to\phi+Ft\), is
\begin{equation}\label{eq:4}
\dot{\phi}_j=\nu\,L_{(1)}(\phi_j)+\frac{\lambda\varepsilon}{2\nu}\phi_j\xi_j,
\end{equation}
with \(1\le j\le N\equiv0\), because periodic boundary conditions
are assumed as usual (the implicit sum convention is not meant in
any of the discrete expressions). The discrete noise \(\xi_j(t)\) is
a Gaussian random variable with zero mean and correlation given by
\begin{equation}\label{dic-noise}
\langle \xi_j(t) \xi_k(t) \rangle= 2 \frac{\delta_{jk}}{a}
\delta(t-t').
\end{equation}

Then, using the discrete version of Eq.\ (\ref{eq:3})
\begin{equation}\label{eq:5}
\phi_j(t)=\exp\left[\frac{\lambda}{2\nu}h_j(t)\right],
\end{equation}
we get
\[\mathrm{e}^{\frac{\lambda}{2\nu}h_j}\frac{\lambda}{2\nu}\dot{h}_j=
\frac{\nu}{a^2}L_{(1)}\left(\mathrm{e}^{\frac{\lambda}{2\nu}h_j}\right)+
\frac{\lambda\varepsilon}{2\nu}\mathrm{e}^{\frac{\lambda}{2\nu}h_j}\xi_j,\]
namely
\[\dot{h}_j=\frac{2\nu^2}{\lambda a^2}\left[\mathrm{e}^{\delta_j^+a}+
\mathrm{e}^{\delta_j^-a}-2\right]+\varepsilon\,\xi_j,\] with
\(\delta_j^\pm\equiv\frac{\lambda}{2\nu}H_j^{j\pm1}\). It is worth
commenting here that this last expression was also pointed out in
\cite{NewmanBray-Discr}, discussing aspects of discretization
instabilities and the relation to the directed polymer problem. We
will further discuss the mapping to the directed polymer problem in
Sec.\ \ref{ssec:2.3} below. By expanding the exponentials up to
terms of order of \(a^2\), and collecting equal powers of \(a\)
(observe that the zero-order contribution vanishes) we retrieve
\begin{equation}\label{eq:6}
\dot{h}_j=\nu\,L_{(1)}(h_j)+\frac{\lambda}{2}Q_{(1)}(h_j)+
\varepsilon\,\xi_j,
\end{equation}
with
\begin{equation}\label{eq:6pp}
Q_{(1)}(h_j)=\frac{1}{2}
\left[\left(H_j^{j+1}\right)^2+\left(H_j^{j-1}\right)^2\right],
\end{equation}
(\(Q\) stands for ``quadratic''). As we see, the first and second
terms on the r.h.s.\ of Eq.\ (\ref{eq:6}) are \emph{necessarily}
related by virtue of Eq.\ (\ref{eq:5}).

\subsection{The general case}\label{ssec:2.2}

A Taylor expansion of \(\phi_{j+l}\) around \(\phi_j\) shows that
the general form of the discrete Laplacian, involving up to the
\(n\)--th nearest neighbors of site \(j\), is of the form
\begin{equation}\label{eq:7}
L_{(n)}(\phi_j)=\frac{\sum_{l=1}^n
b_l\left[\Phi_j^{j+l}+\Phi_j^{j-l}\right]}{a\sum_{l=1}^nl^2b_l},
\end{equation}
where as before, the subscript stands for the number of nearest
neighbors. Since the maximum value for \(n\) is
\(M\equiv(N-1)\backslash2\), where \(\backslash\) denotes integer
division, one may alternatively run the sum up to \(M\) and set
\(b_l=0,\,l=n+1\ldots M\). The remaining \(b_l\), that are otherwise
arbitrary, should be fixed by whatever criterion (below, we shall
use the criterion of maximizing accuracy).

Repeating the steps described above, one obtains
\begin{eqnarray}
L_{(n)}(h_j) & = & \frac{\sum_{l=1}^n
b_l\left[H_j^{j+l}+H_j^{j-l}\right]}{a\sum_{l=1}^nl^2b_l},\label{eq:8}\\
Q_{(n)}(h_j) & = & \frac{\sum_{l=1}^n
b_l\left[\left(H_j^{j+l}\right)^2 +
\left(H_j^{j-l}\right)^2\right]}{2\sum_{l=1}^nl^2b_l}.\label{eq:9}
\end{eqnarray}

\subsection{Few remarks on the directed polymer problem}
\label{ssec:2.3}

We devote this subsection to briefly comment about the mapping of
KPZ onto the directed polymer problem. Such a mapping can be carried
out via the Hopf-Cole transformation \cite{NewmanBray-Discr} and the
resulting linear equation corresponds to Eq.\ (\ref{eq:2}). In order
to employ the usual rules of calculus, here we assume the
Stratonovich interpretation for the multiplicative noise. The
corresponding finite differences scheme is, explicitly,
\begin{equation}\label{findiff}
\dot{\phi}_j=\nu\,L_{(1)}(\phi_j)+\frac{\lambda F}{2\nu}\phi_j+
\frac{\lambda\varepsilon}{2\nu}\phi_j\,\xi_j(t).
\end{equation}
As indicated in Eq.\ (\ref{dic-noise}), the discrete noise
\(\xi_j(t)\) is a Gaussian random variable. The mean value of Eq.\
(\ref{findiff}) is
\begin{equation}
\frac{d\langle\phi_j\rangle}{dt}=\nu\frac{\langle\phi_{j+1}\rangle+
\langle\phi_{j-1}\rangle-2\langle\phi_j\rangle}{a^2}+
\frac{\lambda F}{2\nu}\langle\phi_j\rangle+
\frac{\lambda\varepsilon}{4\nu a}\langle\phi_j\rangle.
\end{equation}
One immediately realizes that the drift of this equation becomes
singular in the continuum limit \(a\to0\), so one has to
renormalize this theory \cite{zinn-justin,Hochberg1,Hochberg2}.
This is done by decomposing the bare parameter into an effective
and a singular component, \(F=F_\mathrm{eff}+F_s\), with
\(F_s=-1/(2a)\). The resulting equation is then
\begin{equation}
\label{meanvalue}
\frac{d\langle\phi_j\rangle}{dt}=\nu\frac{\langle\phi_{j+1}\rangle+
\langle\phi_{j-1}\rangle-2\langle\phi_j\rangle}{a^2}+
\frac{\lambda F_\mathrm{eff}}{2\nu}\langle\phi_j\rangle,
\end{equation}
which is finite, but in which \(F_\mathrm{eff}\) has to be measured
directly from the experiment. Thus the correct interpretation of the
Stratonovich Eq.\ (\ref{findiff}) is the following It\^o equation
\begin{equation}
\dot{\phi}_j=\nu \frac{\phi_{j+1}+\phi_{j-1}-2\phi_j}{a^2}+
\frac{\lambda F_\mathrm{eff}}{2\nu}\phi_j+
\frac{\lambda\varepsilon}{2\nu}\phi_j\xi_j(t).
\end{equation}
In order to measure the effective growth rate, one can solve the
linear Eq.\ (\ref{meanvalue}) to find the globally stable solution
\(\langle\phi_j(t)\rangle=\langle\phi_j(0)\rangle\exp[\lambda
F_\mathrm{eff}\,t/(2\nu)]\) for a spatially homogeneous initial
condition \(\langle\phi_j(0)\rangle=\langle\phi(0)\rangle\). And so
this effective rate can be measured from experimental/numerical data
in the following fashion
\begin{equation}
F_\mathrm{eff}=\frac{2\nu}{\lambda t}
\ln[\langle\phi_j(t)\rangle/\langle\phi_j(0)\rangle],
\end{equation}
or alternatively
\begin{equation}
F_\mathrm{eff}=\frac{2\nu}{\lambda t}
\ln\left\{\langle\exp[\lambda h_j(t)/(2\nu)]\rangle\right\},
\end{equation}
assuming that the initial condition is \(h_j(0)=0\). Applying
Jensen's inequality to this last relation one finds
\begin{equation}
\langle h_j(t)\rangle\ge F_\mathrm{eff}\,t,
\end{equation}
in agreement with what one could directly obtain from the KPZ
equation [Eq.\ (\ref{eq:1})].

\section{Exploiting the deterministic Lyapunov functional}
\label{sec:3}

An important feature of the Hopf--Cole transformation---Eq.\
(\ref{eq:3}) or (\ref{eq:5})---is that it is \emph{local}, i.e.\ it
involves neither spatial nor temporal transformations. Some effects
of this feature are the following
\begin{enumerate}
\item The discrete form of the Laplacian---namely the operator
\(L_{(n)}\)---is the same, regardless of whether it is applied
to \(\phi\) or to \(h\).
\item For a given \(L_{(n)}\) (i.e.\ a given set of
\(b_l=0,\,l=1\ldots n\)), \(Q_{(n)}\) is \emph{determined} by
the Hopf--Cole transformation, Eq.\ (\ref{eq:5}).
\end{enumerate}
In this section, we want to go further with the criterion that the
definitions of the discrete operators should not depend on the
fields on which they are applied.

The \emph{deterministic} part of Eq.\ (\ref{eq:2}), namely the
diffusion term, admits a local Lyapunov functional. In other words,
for \(\varepsilon=0\), Eq.\ (\ref{eq:2}) can be written in the
following variational form
\begin{equation}\label{eq:10}
\partial_t\phi=-\frac{\delta\mathcal{F}[\phi]}{\delta\phi},
\end{equation}
with
\begin{equation}\label{eq:11}
\mathcal{F}[\phi]=\frac{\nu}{2}\int\mathrm{d}x\,(\partial_x\phi)^2.
\end{equation}

The aforementioned criterion dictates the following set of discrete
forms (thus Lyapunov functions, for any finite \(N\)) of Eq.\
(\ref{eq:11})
\begin{equation}\label{eq:12}
\mathcal{F}_{(n)}[\phi]=\frac{1}{2}\nu a \sum_{j=1}^N
Q_{(n)}(\phi_{j}).
\end{equation}
It is a trivial task to verify that
\begin{equation}\label{eq:13}
\nu\,L_{(n)}(\phi_j)=-\frac{1}{a}
\frac{\partial\mathcal{F}_{(n)}[\phi]}{\partial\phi_j}.
\end{equation}
There is no loss of generality in taking \(j=N\equiv0\). If we
rearrange the sum in Eq.\ (\ref{eq:12}) as \(\sum_{j=M+1-N}^M\),
with \(M\equiv(N-1)\backslash2\), then only \(-n\leq j \leq n\) will
contribute in Eq.\ (\ref{eq:13}). Moreover, their contribution is
such that they cancel the factor 1/2 in front of the sum in Eq.\
(\ref{eq:12}). For completeness, let us show the particular
functional form for \(Q_{(1)}(\phi_{j})\)
\begin{equation}\label{eq:12pp}
\mathcal{F}_{(1)}[\phi]=\frac{\nu}{4\,a}\sum_{j=1}^N
\left[(\phi_{j+1}-\phi_j)^2+(\phi_j-\phi_{j-1})^2\right].
\end{equation}

\subsection{Other discrete forms of the KPZ term}\label{ssec:3.1}

Of course, Eq.\ (\ref{eq:13}) does not \emph{uniquely} determine the
Lyapunov function. Expressions other than Eq.\ (\ref{eq:12}) may
yield \(L_{(n)}(\phi_j)\), provided that they contain the right
terms, in the right proportion. Take as an example the proposal of
Refs.\ \cite{LamShin-Discretiz-1,LamShin-Discretiz-2}, coded as
\(N_j^{(1/2)}\) in Ref.\ \cite{Buceta-Discr}
\begin{equation}\label{eq:14}
N^{(1/2)}(\phi_j)=\frac{1}{3}\left[\left(\Phi_j^{j+1}\right)^2
+\left(\Phi_j^{j-1}\right)^2-\Phi_j^{j+1}\Phi_j^{j-1}\right].
\end{equation}
By using \(N^{(1/2)}(\phi_{j'})\) instead of \(Q_{(n)}(\phi_{j'})\)
in Eq.\ (\ref{eq:11}), we obtain
\begin{eqnarray}\label{eq:15}
L_{(2)}^{(1/2)}(\phi_j) & \equiv & \frac{1}{6} \left[2
\left(\Phi_j^{j+1} + \Phi_j^{j-1}\right) + \left(\Phi_j^{j+2} +
\Phi_j^{j-2}\right)\right] \nonumber \\
& = & \frac{1}{6}\left[2L_{(1)}(\phi_j) + \frac{1}{a}
\left(\Phi_j^{j+2} + \Phi_j^{j-2}\right)\right],
\end{eqnarray}
which is an instance of \(L_{(2)}(\phi_j)\), with \(b_1=2\),
\(b_2=1\). As it was shown before, the procedure outlined in Sec.\
\ref{sec:2} will yield \(L_{(2)}^{(1/2)}(h_j)\), together with
\begin{equation}\label{eq:16}
Q_{(2)}^{(1/2)}(h_j)\equiv
\frac{1}{12}\left\{2\left[(H_j^{j+1})^2+(H_j^{j-1})^2\right]
+\left[(H_j^{j+2})^2+(H_j^{j-2})^2\right]\right\}
\end{equation}
and \emph{not} \(N^{(1/2)}(h_j)\). Hence, the proposal of Refs.\
\cite{LamShin-Discretiz-1,LamShin-Discretiz-2} is not consistent: On
one hand, \(N^{(1/2)}(h_j)\) does not correspond with
\(L_{(1)}(h_j)\), as it is used. On the other hand, it does not
correspond with \(L_{(2)}^{(1/2)}(h_j)\) either, as shown.

As stated before, the proposal of Refs.\
\cite{LamShin-Discretiz-1,LamShin-Discretiz-2} belongs to a family
coded in Ref.\ \cite{Buceta-Discr} as \(N_j^{(\gamma)}\),
\(\gamma\in[0,1]\). The choice \(\gamma=1\) yields
\(N^{(1)}(\phi_j)=\frac{1}{4}\left(\Phi_{j-1}^{j+1}\right)^2\). On
the other hand, the choice \(\gamma=0\) yields
\(L_{(2)}^{(0)}(h_j)=L_{(1)}(h_j)\). They all correspond to \(n=2\),
with \(b_1=1-\gamma\), \(b_2=\gamma/2\). The equivalent of Eq.\
(\ref{eq:14}) is now
\begin{equation}\label{eq:17}
N^{(\gamma)}(\phi_j)=\frac{1}{2(1+\gamma)a^2}
\left[\left(\Phi_j^{j+1}\right)^2+\left(\Phi_j^{j-1}\right)^2
-2\gamma\Phi_j^{j+1}\Phi_j^{j-1}\right],
\end{equation}
which \(\forall\gamma\in[0,1]\) yields
\begin{eqnarray}
L_{(2)}^{(\gamma)}(h_j)& \equiv & \frac{1}{1+\gamma}
\left[(1-\gamma)\left(H_j^{j+1}+H_j^{j-1}\right)
+\frac{\gamma}{2}\left(H_j^{j+2}+H_j^{j-2}\right)\right]\nonumber\\
& = & \frac{1}{2(1+\gamma)} \left\{2L_{(1)}(h_j) +
\gamma\left[L_{(1)}(h_{j+1})+L_{(1)}(h_{j-1})\right]\right\}\label{eq:18}
\end{eqnarray}
and
\begin{eqnarray}\label{eq:19}
Q_{(2)}^{(\gamma)}(h_j)\equiv
\frac{1}{2(1+\gamma)}\left\{(1-\gamma)\left[(H_j^{j+1})^2+(H_j^{j-1})^2\right]
+\frac{\gamma}{2}\left[(H_j^{j+2})^2+(H_j^{j-2})^2\right]\right\}.
\end{eqnarray}
However, the accuracy of this discretization is unknown and should
be studied.

\subsection{A more accurate discretization scheme}\label{ssec:3.2}

Again, a Taylor expansion of \(\phi_{j+l}\) around \(\phi_j\) shows
that the \(\mathcal{O}(a^2)\) corrections to \(L_{(n)}\) [applied to
\(h_j\) in Eq.\ (\ref{eq:8})] are of the form
\[\frac{2}{4!}\frac{\sum_{l=1}^nl^4b_l}{\sum_{l=1}^nl^2b_l}\partial_x^4h.\]
Thus, the \(\mathcal{O}(a^2)\) correction to
\(L_{(2)}^{(\gamma)}\) is
\(\frac{1}{12}\frac{1+7\gamma}{1+\gamma}\partial_x^4h\). It
attains its minimum value (\(\frac{1}{12}\partial_x^4h\))
precisely for \(\gamma=0\), namely for \(L_{(1)}\). What is then
the convenience of a more complex prescription for the Laplacian?

A wise criterion for choosing \(b_1\) and \(b_2\) in \(L_{(2)}\)
is making the \(\mathcal{O}(a^2)\) corrections vanish. This yields
the prescription \(b_1=16\), \(b_2=-1\), known to be accurate up
to corrections of \(\mathcal{O}(a^4)\) \cite{AbSt}.

Carrying out the procedure sketched in Sec.\ \ref{ssec:2.1}, we
obtain
\begin{eqnarray}
L_{(2)}(h_j) & \equiv & \frac{4}{3}L_{(1)}(h_j)-\frac{1}{12}
\left(H_j^{j+2}+H_j^{j-2}\right),\label{eq:20}\\
Q_{(2)}(h_j) & \equiv &
\frac{2}{3}\left[(H_j^{j+1})^2+(H_j^{j-1})^2\right]
-\frac{1}{24}\left[(H_j^{j+2})^2+(H_j^{j-2})^2\right].\label{eq:21}
\end{eqnarray}

The \(\mathcal{O}(a^2)\) corrections to \(Q_{(n)}\) are
\[\frac{2}{4!}\frac{\sum_{l=1}^nl^4b_l}{\sum_{l=1}^nl^2b_l}
\left[3\left(\partial_x^2h\right)^2+4\left(\partial_x^3h\right)
(\partial_xh)\right],\] which also vanishes for \(b_1=16\), \(b_2=-1\).

Since this discretization scheme fulfills the consistency
conditions, is accurate up to \(\mathcal{O}(a^4)\) corrections, and
its prescription is not more complex than the ones studied before,
it is obvious that it will be a convenient one to be used when a
higher accuracy in numerical schemes is required. The possibility
that it may also help to control (or at least delay) the numerical
instabilities found in previous works (see for instance
\cite{NewmanBray-Discr,dasgupta} and references therein) will be the
subject of further work.

Let us remind again that, as we already pointed out in the
introduction, these results are formal since the higher order
derivatives of the field are not under control.

\section{Pseudo-spectral discretization}\label{sec:4}

As was indicated in the introduction, a pseudo-spectral spatial
discretization scheme has been recently introduced
\cite{GiaRo01,GiadaGiacomettiRossi,GiaRos,GCL1}. In this section we
show the relation existing between the present analysis and the
indicated pseudo-spectral spatial discretization scheme.

The pseudo-spectral discretization procedure starts by Fourier
expanding the field \(h(x,t)\)
\begin{equation}\label{eq:22}
h(x,t)=\sum_{k=-\infty}^\infty\widehat{h}_k(t)
\exp\left(\mathrm{i}\frac{2\pi}{L}kx\right),
\end{equation}
with
\begin{equation}\label{eq:23}
\widehat{h}_k(t)=\frac{1}{L}\int_0^L\mathrm{d}x\,h(x,t)
\exp\left(-\mathrm{i}\frac{2\pi}{L}kx\right),
\end{equation}
and Eq.\ (\ref{eq:3}) can be rewritten as
\begin{equation}\label{eq:24}
\phi(x,t)=\exp\left[\frac{\lambda}{2\nu}\sum_{k =
-\infty}^\infty\widehat{h}_k(t)\exp\left(\mathrm{i}\frac{2\pi}{L}kx\right)\right].
\end{equation}
Thus Eq.\ (\ref{eq:2}) reads
\begin{equation}\label{eq:25}
\sum_{k=-\infty}^\infty\exp\left(\mathrm{i}\frac{2\pi}{L}kx\right)
\left\{\dot{\widehat{h}}_k(t)+\left(\frac{2\pi}{L}\right)^2k\widehat{h}_k(t)\left[\nu
k+\frac{\lambda}{2}\sum_{k'=-\infty}^\infty
k'\widehat{h}_{k'}(t)\exp\left(\mathrm{i}\frac{2\pi}{L}k'x\right)\right]
-\varepsilon\,\widehat{\xi}_k(t)\right\}=0,
\end{equation}
since \(\xi(x,t)\) is also assumed to be \(L\)--periodic as a
function of \(x\). A \emph{sufficient} condition is that
\begin{equation}\label{eq:26}
\dot{\widehat{h}}_k(t)=-\left(\frac{2\pi}{L}\right)^2k\widehat{h}_k(t)\left[\nu
k+\frac{\lambda}{2}\sum_{k'=-\infty}^\infty
k'\widehat{h}_{k'}(t)\exp\left(\mathrm{i}\frac{2\pi}{L}k'x\right)\right]
+\varepsilon\,\widehat{\xi}_k(t).
\end{equation}
In this context ``discretize'' means to consider only \(N\) Fourier
modes, including \(k=0\). If \(M\equiv(N-1)\backslash2\), then
\begin{equation}\label{eq:27}
\dot{\widehat{h}}_k(t)=-\left(\frac{2\pi}{L}\right)^2k\widehat{h}_k(t)
\left[\nu k+\frac{\lambda}{2}\sum_{k'=M+1-N}^M
k'\widehat{h}_{k'}(t)\exp\left(\mathrm{i}\frac{2\pi}{L}k'x\right)\right]
+\varepsilon\,\widehat{\xi}_k(t).
\end{equation}

As indicated before, it is interesting to connect real-space and
pseudo-spectral discretization approaches. From Eq.\ (\ref{eq:22})
[with \(h(ja,t)\equiv h_j(t)\)] we have
\begin{equation}\label{eq:28}
H_j^{j+l}=\frac{2}{a}\sum_{k=M+1-N}^M\widehat{h}_k(t) \,
\sin\left(\frac{\pi kl}{N}\right) \,
\exp\left[\mathrm{i}\frac{2\pi}{N}k\left(j+\frac{l}{2}\right)\right],
\end{equation}
and
\begin{equation}\label{eq:29}
L_{(M)}(h_j)=\sum_{k=M+1-N}^M\widehat{h}_k(t)\left\{
\frac{2a\sum_{l=1}^Mb_l\left[\cos\left(\frac{2\pi kl}{N}\right)-1\right]}
{a\sum_{l=1}^Ml^2b_l}\right\}\exp\left(\mathrm{i}\frac{2\pi}{N}kj\right)
\end{equation}
By equating this expression to
\[-\sum_{k=M+1-N}^M\widehat{h}_k(t)\left(\frac{2\pi k}{L}\right)^2
\exp\left(\mathrm{i}\frac{2\pi}{N}kj\right),\] we might think of the
pseudo-spectral discretization as a particular real-space
discretization, whose coefficients are the solutions of the linear
system
\begin{equation}\label{eq:29b}
\sum_{l=1}^Mb_l\left[\cos\left( \frac{2\pi kl}{N} \right) - 1 +
\frac{1}{2}\left(\frac{2\pi kl}{N}\right)^2\right]=0.
\end{equation}
This equation is linear and homogeneous, and so it admits the
trivial solution \(b_l=0\:\forall\,l\in\{1,\cdots,M\}\). This
equation expresses in fact the fundamental difference of the
spectral and finite differences discretization: the lattice
spectrum. For the finite difference scheme the discrete Laplacian is
no longer \((2\pi k/L)^2\) (namely twice the spectrum in the
continuum) but \(1-\cos\frac{2\pi kx}{L}\). For \(x=L/2\), already
for \(k=1\) the difference is \(\pi^2/2-2\approx3\). If we equate
instead Eq.\ (\ref{eq:29}) to
\[\sum_{k=M+1-N}^M\widehat{h}_k(t)\left[\cos\left(\frac{2\pi kl} {N} \right)
-1\right] \exp\left(\mathrm{i}\frac{2\pi}{N}kj\right),\] then Eq.\
(\ref{eq:29b}) says nothing new. There is still a complete
arbitrariness in the choice of the coefficients \(b_l\). This
corresponds to the fact that this scheme is nothing but the Fourier
transformed version of the finite differences one.

\section{Galilean Invariance and Fluctuation--Dissipation
Relation}\label{sec:5}

There are two main symmetries associated with the 1D KPZ equation:
the fluctuation--dissipation relation and Galilean invariance. On
one hand the fluctuation--dissipation relation essentially tells us
that the nonlinearity is asymptotically (that is, at long times) not
operative in 1D. On the other hand, Galilean invariance has been
traditionally related to the exact relation among exponents
\(\alpha+z=2\), that holds for all spatial dimensions
\cite{FoNeSt,mhkz89}. However, it is worth remarking that this
interpretation has been recently criticized \cite{BeHo}.

\subsection{Galilean Invariance}

Galilean invariance means that the KPZ equation is invariant under
the transformation
\begin{eqnarray}
x & \to & x -\lambda v\,t,\nonumber\\
h & \to & h + v\,x,\nonumber\\
F & \to F - \frac{\lambda}{2}\,v^2,
\end{eqnarray}
where \(v\) is an arbitrary constant vector field and \(F\) is the
external (constant) driving force. Using the classical
discretization
\begin{equation} \label{discgrad}
\partial_x h\to\frac{1}{2}\,H_{j-1}^{j+1},
\end{equation}
for the (complete) KPZ equation, we find
\begin{equation}\label{compKPZ}
\dot{h}_j=\nu\,L_1+\frac{\lambda}{8}\left(H_{j-1}^{j+1}\right)^2
+F+\xi_j(t).
\end{equation}
One can immediately check that this equation is invariant under the
\emph{discrete} Galilean transformation
\begin{eqnarray} \label{galtrans}
j\,a & \to & j\,a - \lambda v\,t,\nonumber\\
h_j & \to & h_j + v\,j\,a,\nonumber\\
F & \to & F - \frac{\lambda}{2}\,v^2.
\end{eqnarray}
However, Eq.\ (\ref{compKPZ}) has been criticized for its instability
properties, at least when the spatial discretization is not fine
enough \cite{NewmanBray-Discr}. If we use the alternative
discretization
\begin{equation}\label{dkpz}
\dot{h}_j=\nu L_1(h_j)+\frac{\lambda}{4}\left[\left(H_{j}^{j+1}
\right)^2+\left(H_{j-1}^{j}\right)^2\right]+F+\xi_j(t),
\end{equation}
we find that this equation is \emph{not} invariant under the
discrete Galilean transformation. In fact, the transformation \(h\to
h+vja\) yields an excess term which is compatible with the gradient
discretization in Eq.\ (\ref{discgrad}); however, this
discretization does not allow to recover the quadratic term in Eq.\
(\ref{dkpz}), indicating that this finite differences scheme does
not fulfill Galilean invariance. The Hopf--Cole transformed equation
\begin{equation}
\dot{\phi}_j=\nu L_j(\phi)+\frac{\lambda F}{2\nu}\phi_j+
\frac{\lambda}{2\nu}\phi_j\,\xi_j(t),
\end{equation}
\emph{is} Galilean invariant, i.e., it is invariant under the
transformation indicated in Eqs.\ (\ref{galtrans}). Hence, the
nonlinear Hopf-Cole transformation is responsible for the loss of
Galilean invariance. Note that these results are independent of
whether we consider this discretization scheme or a more accurate
one.

Galilean invariance has been always associated with the exactness of
the 1D KPZ exponents, and with a relation that connects the critical
exponents in higher dimensions. If the numerical solution obtained
from a finite differences scheme as Eq.\ (\ref{dkpz}), which is not
Galilean invariant, \emph{yields the well known critical exponents},
that would strongly suggest that Galilean invariance is not a
fundamental symmetry as usually considered.

In Sec.\ \ref{sec:7} we present some numerical results for the
critical exponents using the consistent discretization schemes
indicated in Eqs.\ (\ref{eq:6pp}) and (\ref{eq:21}), and compare
with those found with the standard one. All the cases exhibit the
same critical exponents. Moreover, let us note that the
discretization used in Refs.\
\cite{LamShin-Discretiz-1,LamShin-Discretiz-2}, which also violates
Galilean invariance, yields the same critical exponents too.

When we compare the classical discretization given by Eq.\
(\ref{compKPZ}), that explicitly reads
\begin{equation}\label{aux-01}
\dot{h}_j= \nu \frac{h_{j+1}+h_{j-1}-2h_j}{a^2} +\frac{\lambda}{2}
\left(\frac{h_{j+1}-h_{j-1}}{2a}\right)^2 + F + \xi_j(t).
\end{equation}
with the alternative one in Eq.\ (\ref{dkpz}), that reads
\begin{equation}\label{aux-02}
\dot{h}_j= \nu \frac{h_{j+1}+h_{j-1}-2h_j}{a^2} +\frac{\lambda}{4}
\left[ \left(\frac{h_{j+1}-h_{j}}{a}\right)^2 +
\left(\frac{h_{j}-h_{j-1}}{a}\right)^2 \right] + F + \xi_j(t),
\end{equation}
we find that this second one presents excess fluctuations with
respect to the first. This can be easily seen by means of the
inequality
\begin{equation}
\left(h_{j+1}-h_{j-1} \right)^2=\left(h_{j+1}-h_j+h_j-h_{j-1}
\right)^2 \le 2\left(h_{j+1}-h_{j} \right)^2 + 2\left(h_{j}-h_{j-1}
\right)^2,\nonumber
\end{equation}
which immediately translates into
\begin{equation}
\frac{\lambda}{2} \left(\frac{h_{j+1}-h_{j-1}}{2a}\right)^2 \le
\frac{\lambda}{4} \left[ \left(\frac{h_{j+1}-h_{j}}{a}\right)^2 +
\left(\frac{h_{j}-h_{j-1}}{a}\right)^2 \right],
\end{equation}
where the inequality is strict unless \(h_j=(h_{j+1}+h_{j-1})/2\),
an event which happens with zero probability (note that in 1D and
for long times, the KPZ interface has independent Gaussian
distributed increments, as Brownian motion). This implies that the
excess fluctuations are genuinely present in the interface dynamics.

The excess fluctuations from Eq.\ (\ref{aux-02}) respect to Eq.\
(\ref{aux-01}) can be explicitly computed: the alternative
discretization scheme may be written as
\begin{eqnarray}\label{excess}
\dot{h}_j & = & \nu\frac{h_{j+1}+h_{j-1}-2h_j}{a^2}
+\frac{\lambda}{2}\left\{\frac{h_{j+1}-h_{j-1}}{2a}\right\}^2+\nonumber\\
& & +
\frac{\lambda}{4a^2}\left[\frac{1}{2}h_{j+1}^2+\frac{1}{2}h_{j-1}^2+
2h_j^2-2h_{j+1}h_j-2h_jh_{j-1}+h_{j+1}h_{j-1}\right]\nonumber\\
& & + F + \xi_j(t),
\end{eqnarray}
where the term between curly brackets denotes the Galilean invariant
fluctuations and the term between square brackets denotes the excess
fluctuations. If the excess fluctuations are comparable to the
Galilean fluctuations then there will be a strong violation of
Galilean invariance. If the critical exponents still persist in this
case, that would indicate that Galilean invariance is not such a
fundamental symmetry as usually considered. This will be discussed
in Sec.\ \ref{sec:7}.

\subsection{Fluctuation--dissipation relation: stationary
probability distribution}

As we have already mentioned, together with Galilean invariance, the
fluctuation--dissipation relation is another fundamental symmetry of
the 1D KPZ equation. It is clear that both these symmetries are
recovered when taking the continuum limit on any reasonable
discretization scheme. And thus, an accurate enough partition must
yield suitable results.

The stationary probability distribution for the KPZ problem in 1D is
known to be \cite{HHZ,BarSta}
\[\mathcal{P}_\mathrm{stat}[h]\sim\exp\left\{\frac{\nu}{2\,\varepsilon}
\int\mathrm{d}x\left(\partial_x h\right)^2\right\}.\] For the
simplest discretization scheme in Eq.\ (\ref{eq:13}), we have
\begin{equation}\label{eq:30}
\mathcal{P}_\mathrm{stat}[h]\sim\exp\left\{\frac{\nu}{2\,\varepsilon}
\sum_j\frac{1}{2}\left[(H_j^{j+1})^2+(H_j^{j-1})^2\right]\right\}.
\end{equation}
Inserting this expression into the stationary Fokker--Planck
equation several terms cancel, and the ones surviving can be
expressed as
\begin{equation}\label{eq:31}
\lambda\nu\sum_j\frac{1}{2}\left[(H_j^{j+1})^2+(H_j^{j-1})^2\right]L_j^{(0)}.
\end{equation}
Clearly, the continuous limit of this expression is of the form
\[\lambda\nu\int\mathrm{d}x\left(\partial_x h\right)^2\partial_x^2h,\]
that, as is well known \cite{HHZ}, is identically zero. A numerical
analysis of Eq.\ (\ref{eq:31}) indicates that this expression is
several orders of magnitude smaller than the value of the pdf's
exponent [Eq.\ (\ref{eq:30})], and typically behaves as
\(\mathcal{O}(1/N)\), where \(N\) is the number of spatial points
used in the discretization. Moreover, using expressions with higher
accuracy for the differential operators one gets an even faster
approach to zero. This indicates that the problem with the
fluctuation--dissipation theorem in \(1+1\), discussed in
\cite{LamShin-Discretiz-2,GiadaGiacomettiRossi} can be just
circumvented using more accurate expressions. It is also worth
commenting that, if a consistent discrete scheme is built from the
discrete scheme in \cite{LamShin-Discretiz-2}, it would also violate
the fluctuation--dissipation relation.

\section{On the variational formulation of KPZ}\label{sec:6}

In Ref.\ \cite{Wio}, a variational formulation was introduced for
the KPZ equation. There it was shown that Eq.\ (\ref{eq:1}) can be
written as
\begin{equation}\label{eq:32}
\partial_th(x,t)=-\Gamma(h)
\frac{\delta\mathcal{G}[h]}{\delta h(x,t)}+\varepsilon\,\xi(x,t);
\end{equation}
where (for \(F=0\))
\begin{equation}\label{eq:33}
\mathcal{G}[h]=\int_\Omega\mathrm{e}^{\frac{\lambda}{\nu}h(x,t)}
\frac{\lambda^2}{8\nu}\left[\partial_{x}h(x,t)\right]^2\mathrm{d}x,
\end{equation}
and the function \(\Gamma(h)\) is given by
\[\Gamma(h)=\left(\frac{2\nu}{\lambda}\right)^2
\mathrm{e}^{-\frac{\lambda}{\nu}h}.\]
The way in which the functionals \(\mathcal{F}[\phi]\) and
\(\mathcal{G}[h]\) are related is also shown in Ref.\ \cite{Wio}.
It is also easy to prove that the functional \(\mathcal{G}[h]\)
fulfills the Lyapunov property \(\partial_t\mathcal{G}[h]\leq0\).

According to the previous results, we can write the discrete version
of Eq.\ (\ref{eq:33}) as
\[\mathcal{G}[h]=\frac{\lambda^2}{8\nu}\sum_j\mathrm{e}^{\frac{\lambda}{\nu}
h_j}\frac{1}{2}\left[(H_j^{j+1})^2+(H_j^{j-1})^2\right].\] Now,
introducing this expression into the discrete version of Eq.\
(\ref{eq:32}), and through a simple algebra, we reobtain Eq.\
(\ref{eq:6}). This reinforces our result, and clearly indicates the
need to be consistent when considering a discrete version of the KPZ
equation.

\section{Some numerical results}\label{sec:7}

We present here some results obtained by numerically integrating the
KPZ equation in 1D. Our aim is to compare the standard
discretization scheme [Eq.\ (\ref{aux-01})] with the consistent ones
presented in Eqs.\ (\ref{eq:8}) and (\ref{eq:31}).

To solve Eq.\ (\ref{eq:1}) we discretize \(h(x,t)\) along the
substrate direction \(x\) with lattice spacing \(a = 1\). We employ
a second-order Runge--Kutta algorithm (see e.g.\ \cite{Toral}) with
periodic boundary conditions. Then the equation of motion
\begin{equation}
\dot{h}_j=\nu\,L(h_j)+\frac{\lambda}{2}Q(h_j)+\varepsilon\,\xi_j=
F(h_j)+\varepsilon\,\xi_j
\end{equation}
is integrated according to the recursive relation
\begin{equation}\label{eq.runge_kutta}
h_j(t+\Delta t)=h_j+\frac{\Delta t}{2}(g_1+g_2)+(\Delta t)^{1/2}u_j,
\end{equation}
with
\begin{eqnarray*}
g_1 & = & F(h_j)\\
g_2 & = & F(h_j+\Delta t g_1+(\Delta t)^{1/2}+u_j),
\end{eqnarray*}
where \(u_j\) is a Gaussian random variable.

Without loss of generality, the interface dynamics can be described in
terms of the dimensionless parameter \(\widetilde{\lambda}=
(2\varepsilon/\nu^3)^{1/2}\lambda\). In practice, we set
\(\nu=\varepsilon=1\) and allow \(\lambda\) to vary.

%%%%%%%%%%%%%%%%%%%%%%%%%%%%%%%%%%%%%%%%
\begin{figure}
\begin{center}
\includegraphics[height=10cm]{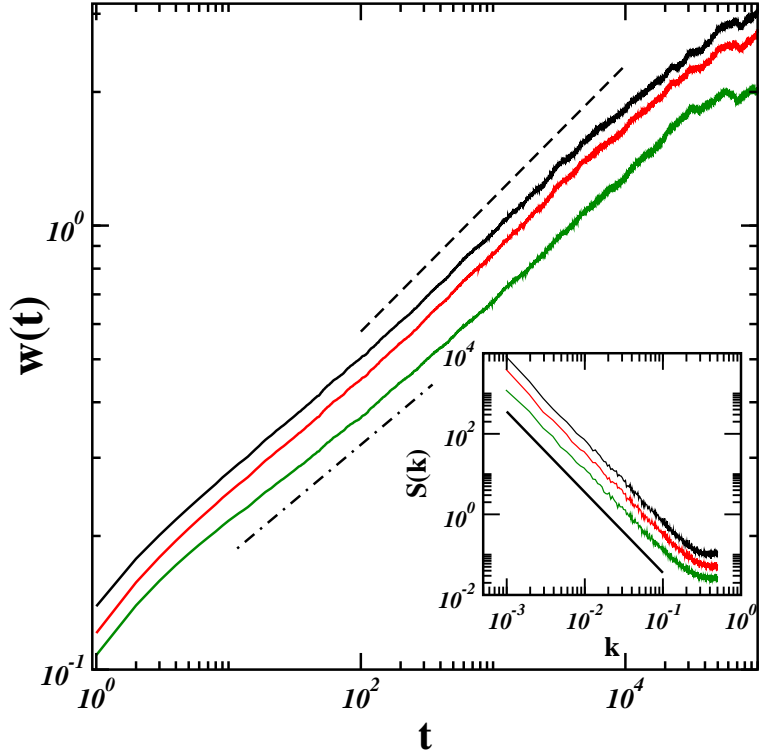}
\caption{Numerical results for the global width and structure factor
(\emph{inset}) averaged over 100 runs in a system of size \(L=1024\)
with \(\lambda=4\). We employ the different discretization schemes
indicated in Eqs.\ (\ref{aux-01}), (\ref{aux-02}) and
(\ref{eq:20},\ref{eq:21}) (from bottom to top). Curves are slightly
shifted vertically for clarity. Lines are plotted as a guide with
exponent \(\beta_\mathrm{KPZ}=0.3\) (\emph{dashed}), \(\beta=0.25\)
(\emph{dot-dashed}), and \(-2\) (\emph{solid, in the inset}). All
discretization schemes are consistent with the KPZ scaling.}
\label{fig.exponents}
\end{center}
\end{figure}
%%%%%%%%%%%%%%%%%%%%%%%%%%%%%%%%%%%%%%%%%

The numerical results show that the interface scaling does not
depend on the discretization scheme. As shown in Fig.\
\ref{fig.exponents}, the dynamics fit into the KPZ universality
class for all the discretization schemes. The global width, that
concerns the fluctuations of the growth height around its mean
value, scales according to the Family--Vicsek Ansatz \cite{BarSta}
as
\begin{equation}\label{eq.anchura_compacta}
W(L,t)=t^\beta f(t/t_x)=t^\beta f(t/L^z),
\end{equation}
where the scaling function \(f(u)\) is defined as
\begin{equation}\label{eq.scaling_anchura}
f(u)\sim\left\{\begin{array}{ll}
\mathrm{const} & \mbox{if }u\ll1,\\
u^{-\beta} & \mbox{otherwise}.\end{array}\right.
\end{equation}

On the other hand, correlations can be analyzed in the reciprocal
space by means of the structure factor
\begin{equation}
\label{eq.structure_factor}
S_k(t)=\langle\widehat{h}_k(t)\widehat{h}_{-k}(t)\rangle,
\end{equation}
where \(\widehat{h}_k(t)\) is as before (see Sec.\ \ref{sec:4}) the
Fourier transform of the interface profile. According to the
previous scaling Ansatz, \(S(k,t)\) scales as \(k^{-(2\alpha+1)}\)
with the roughness exponent \(\alpha\).

We observe that all the discretization schemes are consistent with
the KPZ scaling, with the KPZ exponents \(\alpha=1/2\) and
\(\beta=1/3\). It can also be observed from Fig.\
\ref{fig.exponents} that the crossover from the transient linear
(Edwards--Wilkinson) behavior to the asymptotic nonlinear (KPZ)
behavior appears earlier in both alternative discretization
schemes than in the standard one [Eq.\ (\ref{aux-01})]. This
effect is presumably related to the fact that the nonlinearity of
the alternative schemes always makes a much stronger contribution
to the dynamics than the one in the standard scheme, see Fig.\
\ref{fig.excess_fluct} below. This way, the threshold contribution
from the nonlinearity is received sooner, resulting in an
anticipated departure from the transient linear regime.

In order to analyze the excess of fluctuations that such
discretization schemes present with respect to the standard
one, we extract the Galilean invariant fluctuations from the
quadratic term of the equation of motion. In Fig.\
\ref{fig.excess_fluct} we depict the time dependence of the
different nonlinear contributions for both alternative
discretization schemes. On the left we have the comparison between
the discretization scheme Eq.\ (\ref{aux-01}) and the one in Eq.\
(\ref{aux-02}), while on the right we compare the scheme in Eq.\
(\ref{aux-01}) to the one in Eqs.\ (\ref{eq:20},\ref{eq:21}).

%%%%%%%%%%%%%%%%%%%%%%%%%%%%%%%%%%%%%%%%
\begin{figure}
\begin{center}
\includegraphics[height=6cm]{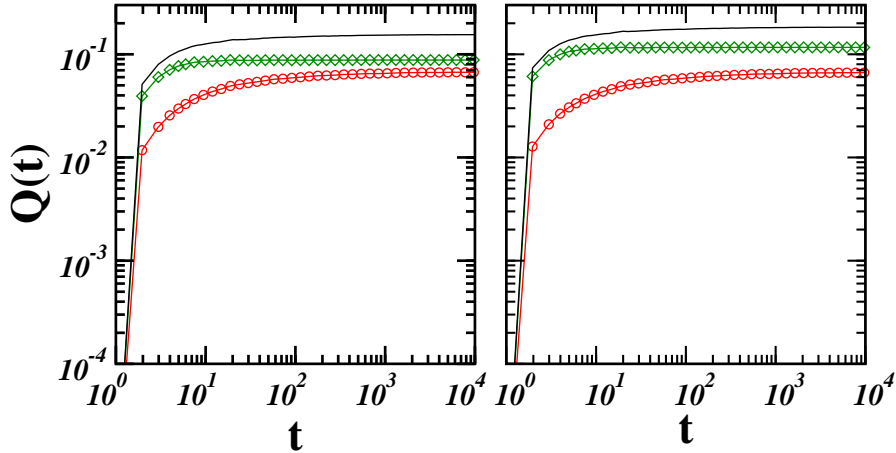}
\caption{Solid line: Time dependence of the nonlinear contribution
in two discretization schemes [Eq.\ (\ref{eq:8}) (left frame) and
Eq.\ (\ref{eq:31}) (right frame)]. We also depict the
Galilean contribution (circles) and the excess of fluctuations
(diamonds) as defined e.g.\ in Eq.\ (\ref{excess}). For both
cases, the excess of fluctuations are comparable with the Galilean
contribution.}\label{fig.excess_fluct}
\end{center}
\end{figure}
%%%%%%%%%%%%%%%%%%%%%%%%%%%%%%%%%%%%%%%%%

The first point to note is the rapid saturation of nonlinearities.
This behavior is consistent with the rapid saturation of local
fluctuations, which behave as \(G(\ell,t)\sim t^{2\beta}\) and
saturate for \(t\gg\ell^z\) \cite{lopez99}.
%Falta breve explicación del significado de \ell y G(\ell,t).
We also observe that for both cases, the excess of fluctuations are
comparable to (or even larger than) the Galilean contribution. As we
pointed out in previous sections, this result, together with the
persistence of KPZ scaling shown in Fig.\ \ref{fig.exponents}, would
imply that Galilean invariance is not such a fundamental symmetry as
usually considered. It is worth remarking that the present results
are not affected by either changing \(L\) or using other algorithms
\cite{delala} to perform the numerical simulations.

\section{Conclusions}\label{sec:8}

The moral from the present analysis is clear: due to the constraint
imposed by the Hopf--Cole transformation [Eqs.\ (\ref{eq:3}) and
(\ref{eq:5})] the discrete forms of the Laplacian and the nonlinear
KPZ term cannot be chosen independently.

Although the present work is focused on the relation between the
diffusion equation with multiplicative noise and the KPZ equation,
the consequences of this analysis are more general. The discrete
versions of any set of related differential equations should be
obtained taking into account the original (or leading) equation and
the transformation rules. It is worth remarking here that a related
analysis was done in \cite{NewmanBray-Discr}, but there the emphasis
was on the study of the strong coupling limit and the mapping onto
the directed polymer problem, without commenting at all about the
consistency among the discrete versions of the differential
operators.

The results discussed here are general; they neither depend on space
dimensionality nor rely on variational representations. Nonetheless,
the recently introduced variational approach for KPZ \cite{Wio}
offers an adequate framework in order to make a consistent
discretization of the KPZ equation.

Regarding the recently introduced pseudo-spectral approach
\cite{GiaRo01,GiaRos,GCL1}, in addition to its known advantages, in
principle, it seems to have the virtue of being ``transparent''
respect to the present problem. In this respect, we have shown the
relation that exists between it and the present analysis. Besides
that, when analyzing inhomogeneous situations where defects or
impurities are present, such methods do not apply and it is again
necessary to resort to real-space discrete form of the differential
operators \cite{delala,ramasco}. Another aspect to consider is
related to the situation in \cite{dasgupta}, where a problem of
numerical instabilities (a computational problem) in discrete growth
models has been tackled by introducing higher order contributions
(changing the physics of the problem!). It is worth indicating that
such an instability does not seem to arise (or at least it arises
latter) in pseudo-spectral treatments of the same problem. Hence,
due to the relation among both formalisms, it seems reasonable to
expect that such instabilities could at least be delayed if a
consistent discretization scheme, together with higher order
discrete operators, is used.

Regarding the two main symmetries associated with the 1D KPZ
equation, the fluctuation--dissipation relation and Galilean
invariance, we have found a couple of relevant results. It is clear
that both these symmetries are recovered when taking the continuum
limit of any reasonable discretization scheme. And thus, an accurate
enough partition must yield suitable results.

The fluctuation--dissipation relation essentially tells us that the
nonlinearity is not operative in 1D and for long times. Our analysis
indicates that the problem with the fluctuation--dissipation theorem
in \(1+1\) can be circumvented by improving the numerical accuracy.
Or this is at least what would happen if the interface were smooth
enough. We are not completely free of surprises coming from the
irregular nature of rough interfaces (as we already mentioned we
expect a H\"older exponent strictly smaller than \(1/2\) for
\(d\)--dimensional KPZ interfaces). In any case, our simulations
have indicated that our strategy of improving the numerical accuracy
yields operative results.

Galilean invariance has been always associated with the exactness of
the 1D KPZ exponents, and with a relation that connects the critical
exponents in higher dimensions. However, it is worth remarking that
this interpretation has been recently criticized \cite{BeHo}. Our
analysis indicates that if the numerical solution obtained with a
finite differences scheme that is not Galilean invariant yields the
well known critical exponents, that would strongly suggests that
Galilean invariance is not a fundamental symmetry as usually
considered. It is worth commenting that the results presented here
for different \emph{consistent} discretization schemes show all the
same critical exponents as the standard one, Eq.\ (\ref{discgrad}).

Here we remark that in the present work we have only emphasized the
existing constraints introduced by the local transformation on the
discrete versions of the differential equations. No attempt is made
here of choosing the most suitable spatial discretization scheme
with regard to a given KPZ feature, nor to present a deep analysis
of results regarding the violation of Galilean invariance. The study
of such aspects, together with the evaluation of the effects of the
relations obtained among the discrete operators on different
relevant quantities as well as other problems will be the subject of
further work.

\begin{acknowledgements}
The authors thank R. Cuerno, H. Fogedby, J.M. L\'opez and M.A.
Rodr\'iguez for fruitful discussions and/or valuable comments,
as well as financial support from the Spanish Government:
Project CGL2007-64387/CLI from MEC (HSW and JAR), Projects
MTM2008-03754 (CE) and FIS2006-12253-C06-04 (MSL) from MICINN.
RRD acknowledges financial support from CONICET and UNMdP of
Argentina. The international collaboration has been facilitated
by AECID, Spain, through Projects A/013666/07 and A/018685/08.
\end{acknowledgements}


\begin{thebibliography}{99}

\bibitem{kpz} M. Kardar, G. Parisi and Y.-C. Zhang, Phys. Rev. Lett.
\textbf{56}, 889 {1986}.

\bibitem{HHZ} T. Halpin-Healy and Y-C. Zhang, Phys. Rep.
\textbf{254}, 215 {1995}.

\bibitem{BarSta} A.-L. Barab\'asi and H. E. Stanley, \textit{Fractal
concepts in surface growth}, (Cambridge U.P., 1995, Cambridge).

\bibitem{Fogedby} H.C. Fogedby and W. Ren, Phys. Rev. E,
\textbf{80}, 041116 (2009).

\bibitem{FoTo} B.M. Forrest and R. Toral, J. Stat. Phys.
\textbf{70}, 703 (1993).

\bibitem{BecCur} M. Beccaria and G. Curci, Phys. Rev. E \textbf{50},
4560 (1994).

\bibitem{MoserWolf-Discr3d} K. Moser and D.E. Wolf, J. Phys. A: Math.
Gen. \textbf{27}, 4049 (1994).

\bibitem{Scalerandi-etal} M. Scalerandi and P. P. Delsanto and S.
Biancotto, Comp. Phys. Comm. \textbf{97}, 185 (1996).

\bibitem{NewmanBray-Discr} T.J. Newman and A.J. Bray, J. Phys. A: Math.
Gen. \textbf{29}, 7917 (1996).

\bibitem{LamShin-Discretiz-1} C.-H. Lam and F.G. Shin, Phys. Rev. E,
\textbf{57}, 6506 (1998).

\bibitem{LamShin-Discretiz-2} C.-H. Lam and F.G. Shin, Phys. Rev. E,
\textbf{58}, 5592 (1998).

\bibitem{Appert} C. Appert, Comp. Phys. Comm., \textbf{121-122}, 363
(1999).

\bibitem{Marinari-etal} E. Marinari, A. Pagnani and G. Parisi, J.
Phys. A: Math. Gen. \textbf{33}, 8181 (2000).

\bibitem{GiaRo01} A. Giacometti and M. Rossi, Phys. Rev. E
\textbf{63}, 046102 (2001).

\bibitem{GiadaGiacomettiRossi} L. Giada, A. Giacometti and M. Rossi,
Phys. Rev. E \textbf{65}, 036134 (2002).

\bibitem{GCL1} R. Gallego, M. Castro and J.M. L\'opez, Phys. Rev. E,
\textbf{76}, 051121 (2007).

\bibitem{Tabei-etal}
S.M.A. Tabei, A. Bahraminasab, A.A. Masoudi, S.S. Mousavi and M.
Reza Rahimi Tabar, Phys. Rev. E \textbf{70}, 031101 (2004).

\bibitem{Reis} F.D.A. Aar\~{a}o Reis, Phys. Rev. E \textbf{72},
032601 (2005).

\bibitem{Buceta-Discr} R. C. Buceta, Phys. Rev. E \textbf{72},
017701 (2005).

\bibitem{Ma-etal-Discr2d} K. Ma, J. Jiang and C.B. Yang, Physica A
\textbf{378}, 194 (2007).

\bibitem{delala} M. S. de la Lama, J.M. L\'opez, J.J. Ramasco and
M.A. Rodr\'iguez, JSTAT P07009 (2009).

\bibitem{FoNeSt} D. Forster, D.R. Nelson and M.J. Stephen, Phys.
Rev. A \textbf{16}, 732 (1977).

\bibitem{mhkz89} E. Medina, T. Hwa, M. Kardar and Y-C. Zhang, Phys.
Rev. A \textbf{39}, 3053 (1989).

\bibitem{BeHo} A. Berera and D. Hochberg, Phys. Rev. Lett.
\textbf{99}, 254501 (2007).

\bibitem{BeHo2} A. Berera and D. Hochberg, Nucl. Phys. B
\textbf{814}, 522 (2009).

\bibitem{Wio} H.S. Wio, Int. J. Bif. Chaos \textbf{19}, 2813 (2009).

\bibitem{nos-KPZ} H.S. Wio, J.A. Revelli, R.R. Deza, C. Escudero and
M.S. de La Lama, submitted to Europhys. Lett. (2009).

\bibitem{schlbr79} A. Schenzle and H. Brand, Phys. Rev. A
\textbf{20}, 1628 (1979).

\bibitem{zinn-justin} J. Zinn-Justin, \textit{Quantum Field Theory
and Critical Phenomena}, (Oxford U.P., Oxford, 2002, 4th ed.).

\bibitem{Hochberg1} D. Hochberg, C. Molina-Par\'{\i}s, J.
P\'erez-Mercader and M. Visser, Physica A \textbf{280}, 437 (2000).

\bibitem{Hochberg2} D. Hochberg, C. Molina-Par\'{\i}s, J.
P\'erez-Mercader and M. Visser, Phys. Lett. A \textbf{278}, 177
(2001).

\bibitem{AbSt} M. Abramowitz and I.A. Stegun, \textit{Handbook of
Mathematical Functions: with Formulas, Graphs, and Mathematical
Tables}, (Dover Publ., New York, 1965).

\bibitem{dasgupta} C. Dasgupta, J.M. Kim, M. Dutta and S. Das Sarma,
Phys. Rev. E \textbf{55}, 2235 (1997).

\bibitem{Toral} M. San Miguel and Ra\'ul Toral, in
\textit{Instabilities and Nonequilibrium Structures VI}, E.
Tirapegui, J. Mart\'{\i}nez-Mardones and R. Tiemann, eds., pgs.
35-130, (Kluwer Ac.Publ., Amsterdam, 2000).

\bibitem{lopez99} J.M. L\'opez, Phys. Rev. Lett. \textbf{83}, 4594
(1999).

\bibitem{GiaRos} A. Giacometti and M. Rossi, Phys. Rev. E
\textbf{62}, 1716 (2000).

\bibitem{ramasco} J.J. Ramasco, J.M. L\'opez and M.A. Rodr\'iguez,
Europhys. Lett. \textbf{76}, 554 (2006).

\end{thebibliography}
\end{document}